\documentclass{aa} \usepackage{epsfig} \usepackage{amssymb}
\usepackage{times} \usepackage{aabib99} 
   \thesaurus{08         % A&A Section 8: Stars,
              (08.01.1;  %  Stars: abundances
               08.16.4;  % Stars: AGB and post-AGB,
               08.05.3;  % Stars: evolution,
               08.09.3;)}% Stars: interiors.
             \title{On the formation of hydrogen-deficient post-AGB
               stars }
             \titlerunning{On the formation of H-deficient post-AGB
               stars}
 
             \author{F.\ Herwig \inst{1}, T.\ Bl\"ocker \inst{2}, N.
               Langer \inst{1} and T.\ Driebe \inst{2} }
             \authorrunning{Herwig et al.\,}
             
             \offprints{F.\ Herwig}
             
             \institute{ Universit\"at Potsdam, Institut f\"ur Physik,
               Astrophysik, Am Neuen Palais 10,
               D-14469 Potsdam\\
               email: fherwig@astro.physik.uni-potsdam.de,
               ntl@astro.physik.uni-potsdam.de \and
               Max-Planck-Institut f\"ur Radioastronomie,
               Auf dem H\"ugel 69, D-53121 Bonn\\
               email: bloecker@speckle.mpifr-bonn.mpg.de,
               driebe@speckle.mpifr-bonn.mpg.de }
             
             \date{Received July 16,1999; accepted August 9, 1999}
\begin{document}
\maketitle

   \begin{abstract}
     We present an evolutionary sequence of a low mass star from the
     Asymptotic Giant Branch (AGB) through its post-AGB stage, during
     which its surface chemical composition changes from hydrogen-rich
     to strongly hydrogen-deficient as consequence of a very late
     thermal pulse, following the so-called {\it born-again scenario}.
     The internal structure and abundance changes during this pulse
     are computed with a numerical method which allows the physically
     consistent calculation of stellar layers where thermonuclear and
     mixing time scale are comparable --- a situation which occurs
     when the helium flash driven convection zone extends to the
     hydrogen-rich surface layers during the pulse peak.  The final
     surface mass fractions are [He/C/O]=[0.38/0.36/0.22], where the
     high oxygen abundance is due to diffusive overshoot employed
     during the AGB evolution. These models are the first to achieve
     general agreement with the surface abundance pattern observed in
     hydrogen-deficient post-AGB stars --- e.g. the PG$\,$1159 stars
     or the WR-type central stars of planetary nebulae ---, confirming
     the born-again scenario with a physically consistent calculation
     and supporting the occurrence of convective overshooting in
     thermally pulsing AGB stars.

   \end{abstract}

%
%________________________________________________________________

\section{Introduction}

Stars on the so called Asymptotic Giant Branch (AGB) have strong
stellar winds, which gradually reduce the mass of the hydrogen-rich
stellar envelope.  When this envelope mass falls below a critical
value, the stars leave the AGB to become post-AGB stars, central stars
of planetary nebulae (CSPNe), and finally white dwarfs.  Post-AGB
stars show a variety of surface abundances \cite{mendez:91}. About
$80\%$ of all CSPNe show a solar-like composition while the remaining
ones are hydrogen-deficient.  Among the latter are Wolf-Rayet type
CSPNe ([WR]-CSPNe) and the extremly hot PG\,1159 stars with typical
surface abundances of [He/C/O]=[0.33/0.50/0.17] (Dreizler and Heber
\cite*{dreizler:98}; see also Koesterke and Hamann
\cite*{koesterke:97b} and references in both papers).
Hydrogen-deficiency is also found in white dwarfs of spectral type DO
\cite{dreizler:96a}.

The origin of the hydrogen-deficiency in post-AGB stars is a
longstanding problem.  Most post-AGB calculations predict a
hydrogen-rich surface composition
\cite{schoenberner:79,schoenberner:83,wood:86,vassiliades:94,bloecker:95b,bloecker:97}.
So far, no post-AGB models reproduced the observed high carbon and
oxygen abundance. The most promising scenario for obtaining a
hydrogen-deficient surface composition envokes a very late thermal
pulse \cite{fujimoto:77,schoenberner:79,iben:83a} --- i.e.\ a pulse
which occurs after the star has already left the AGB --- during which
the pulse driven convection zone can mix hydrogen-free material out to
the stellar surface.  Within this {\it born-again scenario}, Iben and
McDonald \cite*{iben:95b} obtain surface mass fractions of
[He/C/O]=[0.76/0.15/0.01], i.e. their model indeed became strongly
hydrogen-deficient.  Hovever, the large oxygen abundance found in most
H-deficient post-AGB stars could not be reproduced be these, nor by
any other calculation.  These difficulties have posed a strong
limitation to the whole scenario.

In this \emph{Letter} we present a post-AGB model sequence starting
from an AGB model computed with overshoot \cite{herwig:97}, and using
a numerical method of computing nuclear burning and time-dependent
convective mixing simultaneously.

%__________________________________________________________________

\section{Numerical method}
\label{sec:num-method}
The stellar models are based on the stellar evolution code described
by Bl\"ocker \cite*{bloecker:95a}. However, the treatment of the
chemical evolution was entirely replaced by a numerical scheme which
solves the time dependence of the considered nuclear species --- i.e.,
the changes due to thermonuclear reactions and due to mixing --- in
one single step.  This enables us, in contrast to earlier
investigations of very late thermal pulses, to reliably predict the
chemical abundance profiles and the nuclear energy generation rates in
situations where the time scales of nuclear burning and mixing are
comparable.  The abundance change for each isotope at each mesh point
due to diffusive mixing and nuclear processing is given by
\begin{equation}
\label{eq:misch}
\left(\frac{\mathrm{d} \vec{X}_j}{\mathrm{d} t}\right)
= \frac{\partial}{\partial m}\left[\left(4\pi
r^2\rho\right)^2D\frac{\partial \vec{X}_j}{\partial m}\right]
+ \hat{F}_j \cdot \vec{X}_j,
\end{equation}
where $\vec{X}_j$ contains the abundances of all considered isoptopes
at the $j^\mathrm{th}$ mesh point, $\hat{F}_j$ is the nuclear rate
matrix, $D$ is the diffusion coefficient describing the efficiency of
convective mixing, $r$ is the radius, $m$ the mass coordinate and
$\rho$ the density.  This leads to a set of non-linear equations with
$M \cdot N$ unknowns, where M is the number of grid points and $N$ is
the number of isotopes.  In the present calculations, $M$ is of the
order of~2000, and $N=15$ as the main thermonuclear reactions for
hydrogen burning through the pp~chains and the CNO cycle as well as
the main helium burning reactions are included.  The solution is
obtained fully implicit with a Newton-Raphson iteration scheme by
making use of the band-diagonal structure of the problem. The scheme
converges to sufficient precision within about 3 iterations. A coupled
solution of one nuclear reaction at a time and time-dependent mixing,
including also the structure equations, has already been applied by
Eggleton \cite*{eggleton:72}.

%__________________________________________________________________
\section{The AGB starting model}

We start with an AGB model with $\ensuremath{M_{\rm
    ZAMS}}=2\ensuremath{\, {\rm M}_\odot}$ which has been evolved over
22 thermal pulses, including convective overshoot at all convective
boundaries. The treatment and efficiency (f=0.016) of overshoot is the
same as in Herwig et al.\,\cite*{herwig:97}.  In comparison to models
without overshoot the intershell region is much stronger enriched in
carbon and oxygen (mass fractions [He/C/O]=[0.35/0.43/0.19]), which
causes a stronger third dredge-up \cite{herwig:98c}. At the
$16^\mathrm{th}$ thermal pulse (TP) the hydrogen-free core has a mass
of $M_\ensuremath{\mathrm{core}} = 0.573 \ensuremath{\, {\rm M}_\odot}
$ and dredge-up\ starts to operate, leading to a carbon star model at
the last computed TP. At this stage the model star has a total mass of
$M = 1.42 \ensuremath{\, {\rm M}_\odot}$ and
$M_\ensuremath{\mathrm{core}} = 0.604 \ensuremath{\, {\rm M}_\odot} $.
We then artificially increase the mass loss ($\ensuremath{\dot{M}} >
10^{-3}\ensuremath{\, {\rm M}_\odot}/\ensuremath{\, \mathrm{yr}}$) in
order to force the model to leave the AGB at the right phase to
develop a very late thermal pulse.  This procedure is justified for
this exploratory work because it does not affect the nucleosynthesis
and mixing during the very late TP.

%__________________________________________________________________
\section{Evolution through the very late thermal pulse}
\label{sec:post-AGB}

Two cases of the born-again scenario should be distinguished.
Depending on the time when the post-AGB thermal pulse occurs, shell
hydrogen burning may still be active or may already have ceased.

\begin{figure}[hbtp] 
  \resizebox{\hsize}{!}{\includegraphics{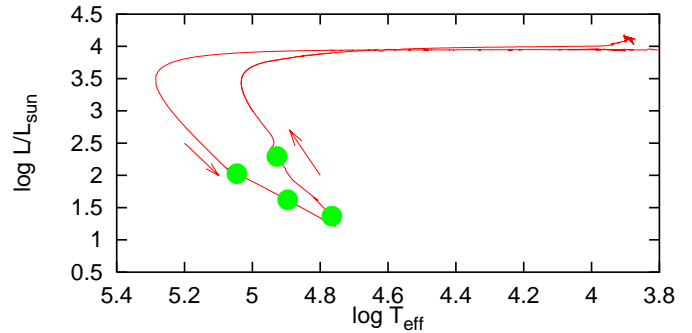}}
\caption{\label{HRD} Track in the HR diagram of our post-AGB model
  during the evolution through a very late TP.  At the first mark
  along the track, the He-flash has already caused a prominent
  convectively unstable region.  At the second mark, the outwards
  growing convection zone has reached the envelope and protons start
  to enter the convective zone (compare Fig.\,\ref{KIPP} and
  Fig.\,\ref{peak-H-flash-a}). At the third mark the hydrogen
  luminosity has reached its peak (see Fig.\,\ref{KIPP} and
  Fig.\,\ref{peak-H-flash}).  The surface composition is hydrogen-free
  beyond the last mark.  }
\end{figure}
\begin{figure}[hbtp] 
  \resizebox{\hsize}{!}{\includegraphics{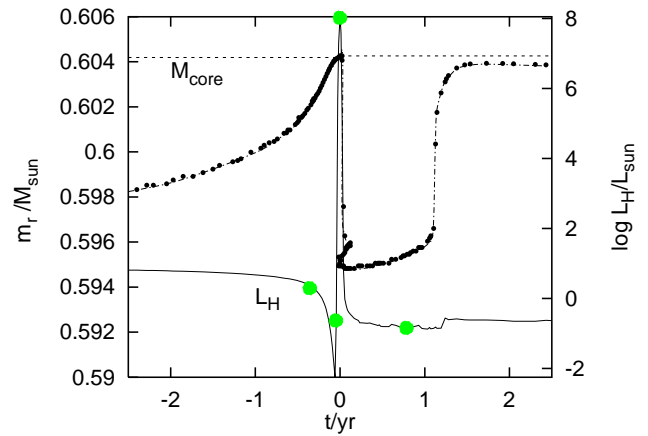}}
\caption{\label{KIPP} Evolution of the top boundary of the convection zone
  (dash-dotted line with filled circle for every second stellar model,
  left scale) as it extends into the hydrogen-rich envelope, and the
  nuclear luminosity due to hydrogen burning (solid line, right
  scale). The four grey dots along the solid line correspond to the
  marks in Fig.\,\ref{HRD}. The dashed line shows the mass coordinate
  of the hydrogen-free core which is for $t>0\ensuremath{\mathrm{\,}}
  \ensuremath{\mathrm{yr}}$ identical with the total stellar mass.  }
\end{figure}
\begin{figure}[hbtp] 
  \resizebox{\hsize}{!}{\includegraphics{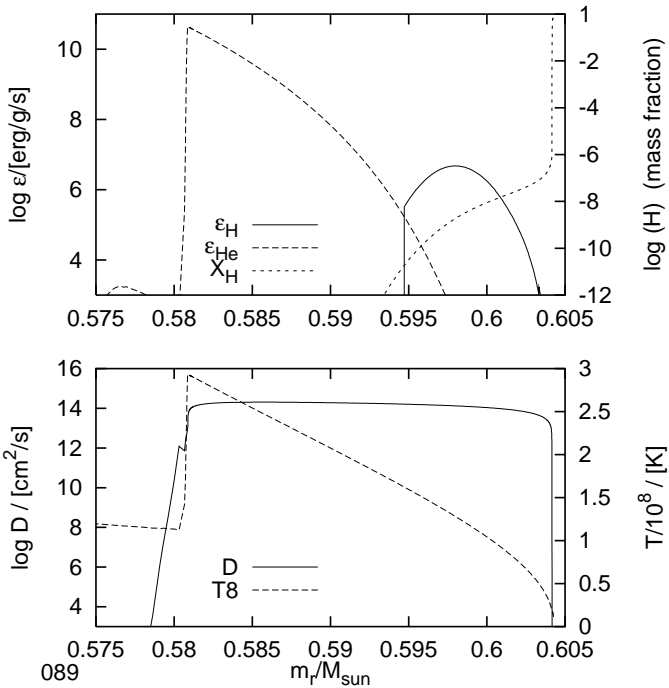}}
\caption{\label{peak-H-flash-a} Internal structure and composition
  at the onset of hydrogen ingestion into the He-flash convection zone
  during the very late TP (second mark in Fig.\,\ref{HRD}).
  \textbf{Top panel:} The nuclear energy generation is dominated by
  processing of helium at the bottom of the He-flash convection zone.
  Energy due to proton capture is released in the upper part of the
  He-flash convection zone (left scale). The hydrogen profile reflects
  the simultaneous nuclear burning and convective mixing (right
  scale). The surface composition at this stage is still
  hydrogen-rich.  \textbf{Bottom panel:} The diffusion coefficient
  (left scale) visualizes the convectively unstable region
  corresponding to the He-flash, which has just reached the lower part
  of the hydrogen-rich envelope.  }
\end{figure}
\begin{figure}[hbtp] 
  \resizebox{\hsize}{!}{\includegraphics{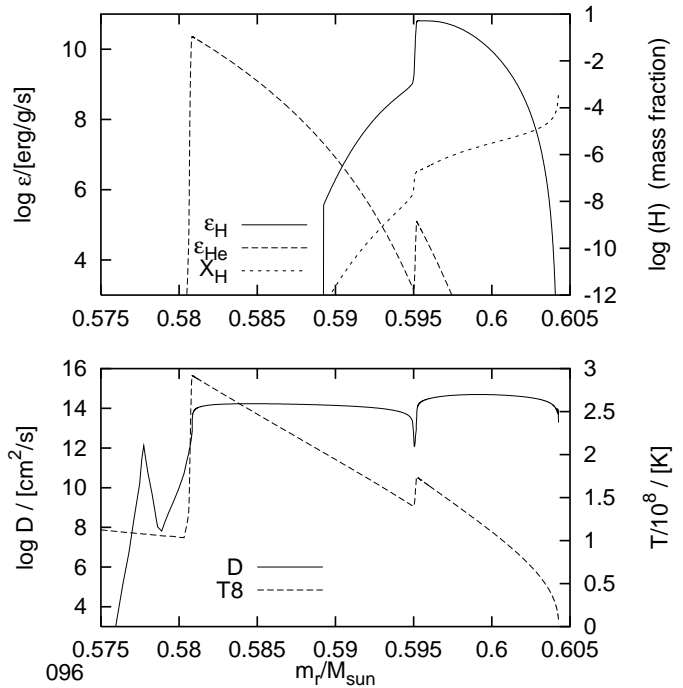}}
\caption{\label{peak-H-flash} Internal structure and composition
  at the time of maximum energy generation due to proton captures
  (third mark in Fig.\,\ref{HRD}). \textbf{Top panel:} Nuclear energy
  generation due to hydrogen burning and helium burning (left scale)
  and hydrogen profile (right scale).  \textbf{Bottom panel:} The
  diffusion coefficient (left scale) shows that the convectively
  unstable region of the thermal pulse is split into two
  ($m_\mathrm{r}=0.595\ensuremath{\, {\rm M}_\odot}$).}
\end{figure}
\begin{figure}[hbtp] 
  \resizebox{\hsize}{!}{\includegraphics{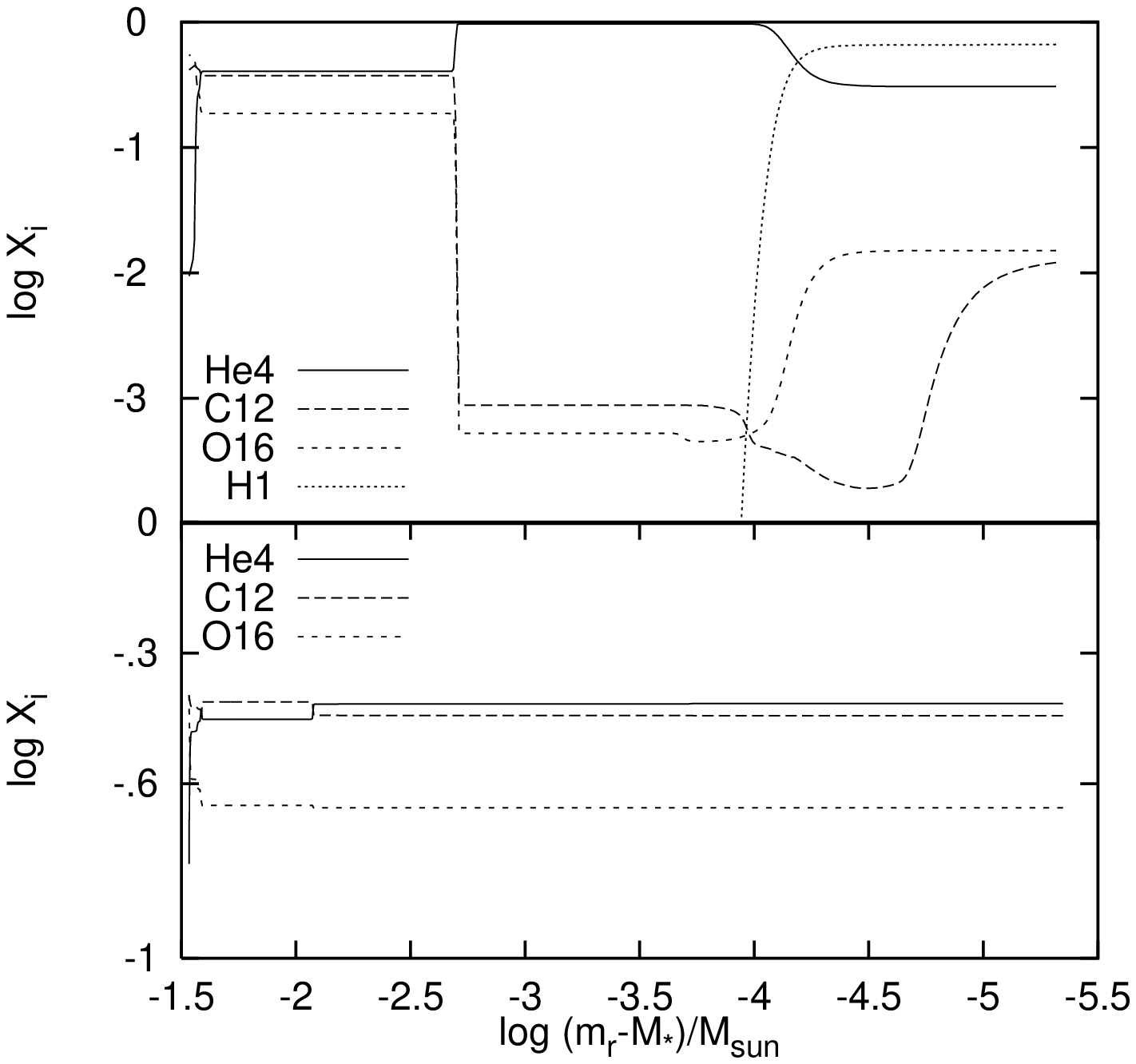}}
\caption{\label{profile} Chemical profiles (mass fraction vs.\ mass
  coordinate) of the upper mass region also covered by
  Fig.\,\ref{peak-H-flash}. \textbf{Top panel:} Profile before the
  convective region of the He-flash has reached the envelope
  corresponding to position of first mark in Fig.\,\ref{HRD}
  \textbf{Bottom panel:} Profile corresponding to position of last
  mark in Fig.\,\ref{HRD} after all hydrogen has been processed (only
  trace amounts of $4 \cdot 10^{-11}\ensuremath{\, {\rm M}_\odot}$ are
  left in the whole star), the abundance at the surface differs only
  slightly from the intershell abundance.}
\end{figure}
In the first case, the He-flash driven convection zone cannot extend
into the hydrogen-rich envelope due to the entropy barrier generated
by the burning shell \cite{iben:76}. In the second case, which is
realized in our model sequence, hydrogen shell burning is extinct and
the star has entered the white dwarf cooling domain (Fig.\,\ref{HRD}).
We designate a TP in this situation as a {\it very late} TP.

As the helium luminosity increases in the course of the He-flash in
our model sequence (first mark in Fig.\,\ref{HRD}), the corresponding
region of convective instability enlarges (Fig.\,\ref{KIPP}). When the
upper convective boundary reaches the mass coordinate where the
hydrogen abundance increases, convective mixing transports protons
downwards into the hot interior (Fig.\,\ref{peak-H-flash-a}).  The
protons are at some point captured by \ensuremath{^{12}\textrm{C}} via
the reaction
$\ensuremath{^{12}\textrm{C}}(p,\gamma)\ensuremath{^{13}\textrm{N}}$.
The peak of the resulting luminosity due to hydrogen burning (see also
Fig.\,\ref{KIPP}) is located at the mass coordinate where the nuclear
time scale equals the mixing time scale ($\simeq \mathrm{one\ hour}$).

The profile of hydrogen in Fig.\,\ref{peak-H-flash-a} and
\ref{peak-H-flash} demonstrates that a correct treatment of
simultaneous burning and convective mixing is essential for this
evolutionary phase. A treatment of convective mixing which does not
include the simultaneous computation of the isotopic abundances
according to the equations of the nuclear network would fail to
predict a correct hydrogen profile.  In particular, such a treatment
would possibly let the protons travel too deep into the convective
region, without considering that they would have been captured already
on the way. Then, the energy generation rate due to proton captures
may be overestimated and not correctly located.

The energy from proton captures is released in the upper part of the
He-flash driven convection zone, which leads to a split (at
$m_\mathrm{r}=0.595\ensuremath{\, {\rm M}_\odot}$) of the convective
region (Fig.\,\ref{peak-H-flash}). The two convective regions are then
connected by the overlapping overshoot extensions, but
Fig.\,\ref{KIPP} shows that the second convective zone is only short
lived since the amount of hydrogen available in the envelope is
quickly consumed.

Figure \ref{peak-H-flash} shows that the hydrogen burning convection
zone extends over $\simeq 10^{-2}\ensuremath{\, {\rm M}_\odot}$ and
reaches from $m_\ensuremath{\mathrm{r}} = 0.595\ensuremath{\, {\rm
    M}_\odot}$ up to the surface of the stellar model. The surface
hydrogen abundance declines rapidly due to mixing and proton captures
in the deeper layers. The period of the largest hydrogen burning
luminosity (shown in Fig.\,\ref{peak-H-flash}) of $\ensuremath{L_{\rm
    H}} \simeq 10^8 \ensuremath{\, {\rm L}_\odot}$ lasts for less than
a week, and the whole episode of convective hydrogen burning is a
matter of about a month. Overall, $5 \cdot 10^{-5}\ensuremath{\, {\rm
    M}_\odot}$ of hydrogen are burnt. At peak hydrogen luminosity the
hydrogen mass fraction at the surface is $3.4 \cdot 10^{-4}$ and the
total amount of hydrogen still present in the star is
$\ensuremath{M_{\rm H}} = 7.8 \cdot 10^{-6}\ensuremath{\, {\rm
    M}_\odot}$. Thus, in this sequence the star is already
hydrogen-deficient before it returns to the AGB domain in the HRD.

Figure \ref{profile} shows abundance profiles before and after the
mixing and burning event due to the very late TP. While the star still
shows the typical hydrogen-rich AGB abundance pattern before the
convective region has reached into the envelope (top panel,
Fig.\,\ref{profile}), the mixing during the convective hydrogen
burning leads to a hydrogen-free surface with
[He/C/O]=[0.38/0.36/0.22] and a mass fraction of $3.5\%$ of neon. The
step in the abundances of \ensuremath{^{4}\textrm{He}},
\ensuremath{^{12}\textrm{C}} and \ensuremath{^{16}\textrm{O}} at
$0.596\ensuremath{\, {\rm M}_\odot}$ (lower panel) corresponds to the
split of the convective region due to hydrogen burning. While the
hydrogen burning leads not to a significant abundance changes for the
major isotopes (only $5 \cdot 10^{-5}\ensuremath{\, {\rm M}_\odot}$ of
hydrogen are processed), helium burning continues to process helium at
the bottom of the He-flash convective zone.  The final surface
abundances are very similar to the intershell abundances during the
thermal pulse.

After most of the hydrogen is burnt, the corresponding upper
convection zone disappears when the local luminosity drops. It takes
about one year until the He-flash convection zone has recovered to its
original extent (Fig.\,\ref{KIPP}). The star then follows the
evolution as known from the born-again scenario (for a recent account
on this scenario see Bl\"ocker and Sch\"onberner,
1997\nocite{bloecker:97}).  Energetically, the return into the AGB
domain is almost exclusively driven by the energy release due to
helium burning, which exceeds the additional supply of energy from
hydrogen burning by orders of magnitude.

%__________________________________________________________________
\section{Conclusions}
\label{sec:conclusions}
Using a numerical method to treat nuclear burning and mixing
simultaneously in stellar evolution calculations, which allows a
reliable and robust modelling of very late thermal pulses, we have
shown that the general surface abundance pattern observed in
hydrogen-deficient post-AGB stars can be explained within the
born-again scenario.  Our new post-AGB sequence shows that due to the
energy generation and convective mixing during a very late thermal
pulse a born-again star forms which displays its previous intershell
abundance at the surface.

We have based the calculation on an AGB model sequence computed with
overshoot, which shows a high carbon and oxygen intershell abundance.
Thus, the fact that the abundance pattern of our post-AGB model after
the thermal pulse agrees with the observation of hydrogen-deficient
post-AGB stars like PG\,1159 and [WC]-CSPNe strongly supports the
assumption of extra mixing beyond the convective boundary of the
He-convection zone in AGB stars.  We conclude that the very late
thermal pulses can indeed be identified as one cause for the
hydrogen-deficiency in post-AGB stars.

However, we note that not all H-deficient post-AGB stars are
completely free of hydrogen \cite{leuenhagen:98}, as predicted by our
model. Other possibilities than the born-again scenario to achieve
H-deficiency might also exist \cite{tylenda:96,waters:98}.  Whether
post-AGB models which are not entirely hydrogen-free can be obtained
within this scenario requires a study of the variation of the late
thermal puls with the inter-pulse phase at which the star leaves the
AGB \cite{iben:84}, and possibly the consideration of other mixing
processes, e.g. due to rotational effects \cite{langer:99}, which has
to be left to future investigations.

\begin{acknowledgements}
  We are grateful to W.-R.~Hamann and L.~Koesterke for many useful
  discussions.  This work has been supported by the \emph{Deut\-sche
    For\-schungs\-ge\-mein\-schaft} through grant La\,587/16.
\end{acknowledgements}
\bibliography{astro}

\begin{thebibliography}{}

\bibitem[\protect\astroncite{Bl\"ocker}{1995b}]{bloecker:95a}
Bl\"ocker, T., 1995b,
\newblock {A\&A} {297}, 727

\bibitem[\protect\astroncite{Bl\"ocker}{1995a}]{bloecker:95b}
Bl\"ocker, T., 1995a,
\newblock {A\&A} {299}, 755

\bibitem[\protect\astroncite{Bl\"ocker and Sch\"onberner}{1997}]{bloecker:97}
Bl\"ocker, T. and Sch\"onberner, D., 1997,
\newblock {A\&A} {324}, 991

\bibitem[\protect\astroncite{Dreizler and Heber}{1998}]{dreizler:98}
Dreizler, S. and Heber, U., 1998,
\newblock {A\&A} {334}, 618

\bibitem[\protect\astroncite{Dreizler and Werner}{1996}]{dreizler:96a}
Dreizler, S. and Werner, K., 1996,
\newblock in C.~S. Jeffery and U. Heber (eds.), {Hydrogen-Deficient Stars},
  Vol.~96, p. 281, ASP Conf.\ Ser.

\bibitem[\protect\astroncite{Eggleton}{1972}]{eggleton:72}
Eggleton, P.~P., 1972,
\newblock {MNRAS} {156}, 361

\bibitem[\protect\astroncite{Fujimoto}{1977}]{fujimoto:77}
Fujimoto, M.~Y., 1977,
\newblock {PASJ} {29}, 331

\bibitem[\protect\astroncite{Herwig et~al.}{1999}]{herwig:98c}
Herwig, F., Bl\"ocker, T., and Sch\"onberner, D., 1999,
\newblock in T.~L. Bertre, A. Lebre, and C. Waelkens (eds.), {AGB Stars},
  p.~41, PASP

\bibitem[\protect\astroncite{Herwig et~al.}{1997}]{herwig:97}
Herwig, F., Bl\"ocker, T., Sch\"onberner, D., and {El Eid}, M.~F., 1997,
\newblock {A\&A} {324}, L81

\bibitem[\protect\astroncite{Iben}{1976}]{iben:76}
Iben, Jr., I., 1976,
\newblock {ApJ} {208}, 165

\bibitem[\protect\astroncite{Iben}{1984}]{iben:84}
Iben, Jr., I., 1984,
\newblock {ApJ} {277}, 333

\bibitem[\protect\astroncite{Iben et~al.}{1983}]{iben:83a}
Iben, Jr., I., Kaler, J.~B., Truran, J.~W., and Renzini, A., 1983,
\newblock {ApJ} {264}, 605

\bibitem[\protect\astroncite{Iben and McDonald}{1995}]{iben:95b}
Iben, Jr., I. and McDonald, J., 1995,
\newblock in D. Koester and K. Werner (eds.), {White Dwarfs}, No. 443 in LNP,
  p.~48, Springer, Heidelberg

\bibitem[\protect\astroncite{{Koesterke} and {Hamann}}{1997}]{koesterke:97b}
{Koesterke}, L. and {Hamann}, W.~R., 1997,
\newblock {A\&A} {320}, 91

\bibitem[\protect\astroncite{Langer et~al.}{1999}]{langer:99}
Langer, N., Heger, A., Wellstein, S., and Herwig, F., 1999,
\newblock {A\&A} {346}, L37

\bibitem[\protect\astroncite{Leuenhagen and Hamann}{1998}]{leuenhagen:98}
Leuenhagen, U. and Hamann, W.-R., 1998,
\newblock {A\&A} {330}, 265

\bibitem[\protect\astroncite{M\'endez}{1991}]{mendez:91}
M\'endez, R.~H., 1991,
\newblock in G. Michaud and A. Tutukov (eds.), {Evolution of Stars:The
  Photospheric Abundance Connection}, p. 375

\bibitem[\protect\astroncite{Sch\"onberner}{1979}]{schoenberner:79}
Sch\"onberner, D., 1979,
\newblock {A\&A} {79}, 108

\bibitem[\protect\astroncite{Sch\"onberner}{1983}]{schoenberner:83}
Sch\"onberner, D., 1983,
\newblock {ApJ} {708}, 272

\bibitem[\protect\astroncite{Tylenda}{1996}]{tylenda:96}
Tylenda, R., 1996,
\newblock in C.~S. Jeffery and U. Heber (eds.), {Hydrogen-Deficient Stars},
  Vol.~96, p. 101, ASP Conf.\ Ser.

\bibitem[\protect\astroncite{Vassiliadis and Wood}{1994}]{vassiliades:94}
Vassiliadis, E. and Wood, P., 1994,
\newblock {ApJS} {92}, 125

\bibitem[\protect\astroncite{Waters et~al.}{1998}]{waters:98}
Waters, L. B. F.~M., Beintema, D.~A., Zijlstra, A.~A., and et~al., 1998,
\newblock {A\&A} {331}, L61

\bibitem[\protect\astroncite{Wood and Faulkner}{1986}]{wood:86}
Wood, P.~R. and Faulkner, D.~J., 1986,
\newblock {ApJ} {307}, 659

\end{thebibliography}
\end{document}